\documentstyle[epsf,aps,multicol]{revtex}
\begin{document}
 %
 %
 %
\let\makelabel=\descriptionlabel
\draft

\title{Stabilization/destabilization of cell membranes by multivalent 
ions: Implications for membrane fusion and division}
\author{Bae-Yeun Ha } 
\address{ Department of Physics, Simon Fraser University, Burnaby, 
B.C., Canada, V5A 1S6}   
\maketitle

\begin{abstract}
We propose a mechanism for the stabilization/destabilization of cell 
membranes by multivalent ions with an emphasis on its implications for 
the division and fusion of cells.  We find that multivalent cations 
preferentially adsorbed onto a membrane {\it dramatically} changes 
the membrane 
stability.  They not only reduce the surface charge 
density of the membrane but also induce a repulsive barrier to pore growth.  
While both of these effects lead to enhanced membrane stability against 
vesiculation and pore growth, 
the repulsive barrier arises from correlated fluctuations of the adsorbed 
cations and favors closure of a pore.  Finally, the addition  
of a small amount of multivalent anions can reverse the membrane 
stabilization, providing an effective way to regulate membrane 
stability.  


\end{abstract}


  \begin{multicols}{2}

\narrowtext

Lipid bilayer membranes are resistant to rupture, primarily serving 
as a barrier to the leakage of the cell's contents, while also being 
dynamic structures that undergo various topological transitions.  
The capability of living cells to regulate the stability of their 
bounding membranes is crucial to their maintenance and 
reproduction~\cite{albert}.  Membrane stability 
against rupture changes most dramatically during 
cell division and fusion. 
The precise mechanism for achieving this complex task in living cells is 
complicated by various membrane-associated/bounded proteins~\cite{albert} and is not 
yet clear.  Numerous studies, however, suggest that  
membrane stability is influenced by several 
factors such as the ionic strength, external fields, and thermal 
fluctuations~\cite{neumann,chang,steck,lieber,lew,helfrich,zhelev,BB,nelson,sung,boal}.
For example, red blood cells can be converted 
into vesicles by osmotic lysis in a solution of low 
ionic strength lacking  multivalent cations~\cite{albert,steck,lieber,lew}.
The presence of divalent cations, however, prevents this 
vesiculation~\cite{steck,lieber,lew}. 
In fact, a number of experiments~\cite{steck,lieber,lew} have 
unambiguously demonstrated that the stability of red cell membranes against 
vesiculation can be greatly enhanced by multivalent cations.  
Despite this, a consistent theoretical description of this phenomenon has 
so far been lacking.       

The strong valency dependency of membrane 
stability~\cite{steck,lieber,lew} motivated this 
work.  Not only can osmotic lysis lead to vesiculation, but it can also create 
large pores in the cell membranes that subsequently contract to a size 
that is controlled by the ionic strength. Pore closure can be 
stimulated by cations, and remarkably the rate of pore closure {\it strongly} 
depends on the valency of cations~\cite{lieber}; ${\rm Ca}^{2+}€$ is 
roughly 60 times as potent on a molar basis as ${\rm Na}^{+}€$.
The potency of divalent cations, which essentially prevents 
vesiculation, was first demonstrated experimentally three decades 
ago~\cite{steck}, but its has yet to be examined theoretically.  Here
we propose a theoretical mechanism to explain this phenomenon.  We find that multivalent counterions 
adsorbed onto charged membranes {\it dramatically} enhance 
the membrane  stability through two effects.  First, they can significantly reduce the strength 
of the electrostatic repulsion between backbone charges on the membrane, 
which enhances the membrane stability.  Second, they induce a repulsive 
barrier to pore growth.  The repulsive barrier originates from the 
correlated fluctuations of adsorbed counterions and favors 
closure of a pore, further stabilizing the membranes against vesiculation 
and pore growth.  
Upon adding a small content of multivalent anions, the adsorbed 
cations are released into solution, 
thus reversing the membrane stabilization.  Adsorption/desorption of 
multivalent cations provide an 
effective way to regulate the membrane stability.        

\begin{figure}
   \par\columnwidth20.5pc
   \hsize\columnwidth\global\linewidth\columnwidth
   \displaywidth\columnwidth
\epsfxsize=2.6truein
\centerline{\epsfbox{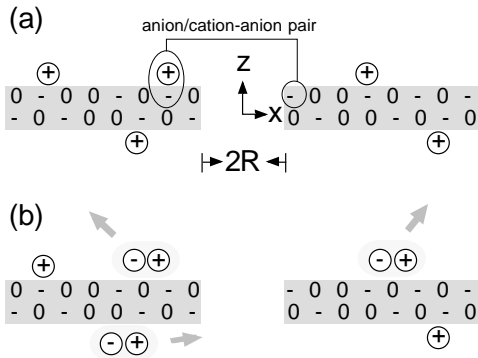}}
\caption{(a) Schematic view of a charged membrane with a pore of a 
radius $R$.  Charged and neutral lipids are denoted by $-$ and $0$, 
respectively, while the adsorbed multivalent cations are denoted by encircled 
$\mbox{\boldmath $+$}$'s.  A pair of attracting lipid and 
lipid/cation is also shown that stabilizes the membrane against pore 
growth.  
(b) Illustration of multivalent anions (encircled 
$\mbox{\boldmath $-$}$'s) forming pairs with 
the adsorbed multivalent cations, then leaving into solution.  Depletion of 
the adsorbed layer destabilizes the membrane.          
 \label{model.fig}}
 \end{figure}

The model we consider here is a thin flat membrane~\cite{remark}  in 
the $xy$-plane, in the 
presence of a monovalent (1:1) salt such as NaCl, and in the presence 
or absence 
of  $Z_{+}€$-valent $(Z_{+}€:1)$ salts, such as ${\rm CaCl}_{2}$.  Each side of 
the membrane is assumed to be negatively charged with constant charge density 
$-e \sigma_{0}€$ and attracts 
ions of the opposite charge, as schematically shown in Fig.~\ref{model.fig}(a).  
For simplicity, we consider the case of a 
single circular pore of radius $R$, already formed in the membrane 
by, for example, osmotic stress.  The stability of the membrane against 
rupture can be quantified in terms of a line tension, {\i.e.}, the energetic 
penalty for creating a pore per unit length.  The electrostatic 
repulsion between charges on the membrane favors pore 
formation~\cite{BB}, but the hydrophobic effect tends to close the pore.                 
If mobile ions are treated as screening objects 
that simply reduce the electrostatic repulsion 
between charges on the membrane via Debye screening, then the electrostatic 
contribution to the line tension $\gamma_{DH}€$ can be estimated 
using Debye-H\"uckel (DH) theory~\cite{BB}: ${\cal \gamma}_{DH}€ \sim - e^{2} 
\sigma_{0}^{2} \kappa^{-1} 
\epsilon^{-1} R$, if $R< \kappa^{-1}€$ and ${\cal \gamma}_{DH}€ 
\sim - e^{2}€\sigma_{0}^{2} \kappa^{-2} \epsilon^{-1}$, if $R > \kappa^{-1}€$, 
where $\kappa^{-1}$ is the screening length and $\epsilon$ is the 
dielectric constant of the solvent.  Obviously, the electrostatic 
repulsion favors creation and expansion of a pore, {\i.e.}, 
$\gamma_{DH} <0$.

Charged membranes are, however, capable of adsorbing counterions of 
the opposite 
charge (See Fig.~\ref{model.fig}(a))~\cite{Alexander,jacob}, reducing 
the surface charge density of the membrane~\cite{Alexander}.  
The magnitude of the reduced, renormalized charge density can be 
estimated by equating the chemical potentials of the ``free'' and 
``condensed'' counterions, {\i.e.}, those adsorbed onto the membrane surface.  
In the following descriptions, the subscripts $i=1$ and $2$ refer to 
the monovalent and multivalent counterions, respectively.  
If $ \sigma_{i}$ is the number density of condensed 
counterions, then the effective (renormalized) surface charge density on the 
membrane is $-e \sigma^{\ast}€ = -e (\sigma_{0} -\sigma_{1} -Z_{+}€ 
\sigma_{2}) $.  The chemical potential of free 
counterions is mainly associated with the configurational entropy of 
mixing: $\mu_{i}^{free}€ \sim k_{B}T \ln \left( n_{i} a_{i}^{3} \right)$, where $n_{i}$ 
and $a_{i}$ are the concentration and size of counterions, respectively.      
On the other hand, the chemical potential of the condensed counterions 
arises from electrostatic interactions and the entropic penalty 
for condensation; $\mu_{i}^{cond}  \sim 
-k_{B}€T (Z_{+}€ \ell_{B} \sigma^{\ast} \sqrt{S}) + \ln \left( \sigma_{i} 
a_{i}^{2} \right)$ 
if $\kappa^{-1} > \sqrt{S},  $where $S$ is the area of 
the membrane, and 
$\mu_{i}^{cond} \sim -k_{B}€T (Z_{+}€ \ell_{B} \sigma^{\ast} 
\kappa^{-1})+\ln \left( \sigma_{i} 
a_{i}^{2} \right) $ otherwise.  The equilibrium values of $\sigma_{i}€$ 
can then be obtained by requiring 
$\mu_{i}^{free}€=\mu_{i}^{cond}€$.  

The condensed counterions do not simply renormalize the membrane 
charge density but also give rise to charge fluctuations in the plane 
of the membrane surface that tend to be correlated with each 
other~\cite{ha.bundle,pincus}.    
A typical attracting pair of a lipid and lipid/cation is illustrated 
in Fig.~\ref{model.fig}.     
Creation of a pore makes charges at the edge less efficiently correlated and  
 is discouraged by the charge correlation effects.                              
Computation of the charge correlation contribution to the pore free 
energy is highly involved, partly because the charge 
fluctuation interactions are not pairwise 
additive~\cite{ha.bundle,parsegian}. 
This is complicated by yet another factor: the specific geometry of 
the membrane with a pore.  The electrostatic effects at the 
meanfield level, suppressing both adsorption and charge 
fluctuations, has only recently been addressed~\cite{BB}.  To study 
the effects of charge fluctuations on the membrane stability, we 
take the continuum limit and incorporate the in-plane charge 
fluctuations at the Gaussian level as in previous 
cases~\cite{ha.bundle,pincus}.  The resulting 
charge fluctuation contribution to the 
pore free energy, {\i.e.}, the change in 
the charge fluctuation free energy by creating a pore, is formally given by
\begin{equation}
\label{fpore}
{\Delta {\cal F}_{pore} \over k_{B}€T} = \mbox{$\frac{1}{2}$}  \ln 
\det \left[ 
1+ (Q- {\cal Q})  {\cal Q}^{-1} \right] 
.\end{equation}       
Here, the matrix $Q$ is defined by the matrix elements
\begin{equation}
Q_{{\bf x}_{\bot} {\bf x}_{\bot}'}=  1+ \ell_{B} \sigma_{cc} \zeta_{{\bf x}_{\bot} {\bf 
x}_{\bot}'}
 {{\rm e}^{- \kappa | {\bf x}_{\bot}-{\bf x}_{\bot}' |}
\over |{\bf x}_{\bot}-{\bf x}_{\bot}'|} 
,\end{equation}   
where ${\bf x}_{\bot} =(x,y)$, 
$\ell_{B}=e^{2}€/\epsilon k_{B}€T$ is the Bjerrum length, 
$\sigma_{cc}€\equiv \sigma_{1}€+Z_{+} \sigma_{2}$, 
$\kappa^{2} = 8 \pi \ell_{B} I$, $I$ is the ionic strength of the 
solution, and ${\cal Q}\equiv \lim_{R \rightarrow 0} Q$.   
Finally $\zeta_{{\bf x}_{\bot} {\bf 
x}_{\bot}'}=1$, if ${\bf x}_{\bot}$ and  ${\bf 
x}_{\bot}'$ are on the membrane and is 0 otherwise.   In the case of $S >  
\kappa^{-1}€$, as is the case for red cell experiments~\cite{steck,lieber}, 
${\cal F}_{pore}€$ in Eq.~(\ref{fpore}) can be calculated without making 
further approximations.  This follows from the fact that $\delta \equiv 
(Q-{\cal Q}) {\cal Q}^{-1} \sim S^{-1}$ and ${\cal F}_{pore}€$ can then be 
expanded in powers of $\delta$.  In the case $S > \kappa^{-1}$, we 
can take the limit $S \rightarrow \infty$ without introducing 
any appreciable error.  In this limit, only the leading term survives in the expansion. 

To calculate the free energy in Eq.~(\ref{fpore}), it proves useful to 
Fourier transform it from ${\bf x}_{\bot}€$ to ${\bf k}_{\bot}€$.  We find 
\begin{eqnarray}
\label{fpore'}
&&\quad {\Delta {\cal F}_{pore} \over k_{B}€T} \simeq {1 \over 2 \lambda_{cc}€} 
\left[ \int\int_{R = 0} 
-\int\int_{R>0} \right] r dr r' dr'  \nonumber \\
&& \times \int_{0}^{2 \pi} d \theta  
\int_{0}^{\infty} {k_{\bot} d k_{\bot} \over 1+ \lambda_{cc}€ 
\sqrt{k_{\bot}^{2}€+\kappa^{2}€}} 
\cdot { {\rm e}^{- \kappa 
             \sqrt{r^{2}+r'^{2}€-2 r r' \cos \theta } }€
              \over 
    \sqrt{r^{2}€+r'^{2}€-2 r r' \cos \theta } } \nonumber \\
    && \qquad \times J_{0}(k_{\bot} \sqrt{r^{2}€+r'^{2}€-2 r r' \cos \theta })
\end{eqnarray}
where $\lambda_{cc} \equiv 1/2 \pi \ell_{B} \sigma_{cc}€$ and 
$J_{0}(x)$ is the zeroth-order Bessel function of the first 
kind.   
In the case of $\kappa R \gg 1$, the charge-fluctuation line 
tension, {\i.e.}, $\Delta \gamma =\Delta {\cal F}_{pore}/2 \pi R$,  
shows two distinct scaling behaviors: $\Delta \gamma \sim \lambda_{cc}^{-1}€\ln (1/ \lambda_{cc}€ 
 \kappa)$ for small $\lambda_{cc} \kappa$ and $\Delta \gamma \sim 
 \lambda_{cc}^{-2}  \kappa^{-1}$ for large $\lambda_{cc} \kappa$.    

\begin{figure}
   \par\columnwidth20.5pc
   \hsize\columnwidth\global\linewidth\columnwidth
   \displaywidth\columnwidth
   \epsfxsize=3.12truein
   \centerline{\epsfbox{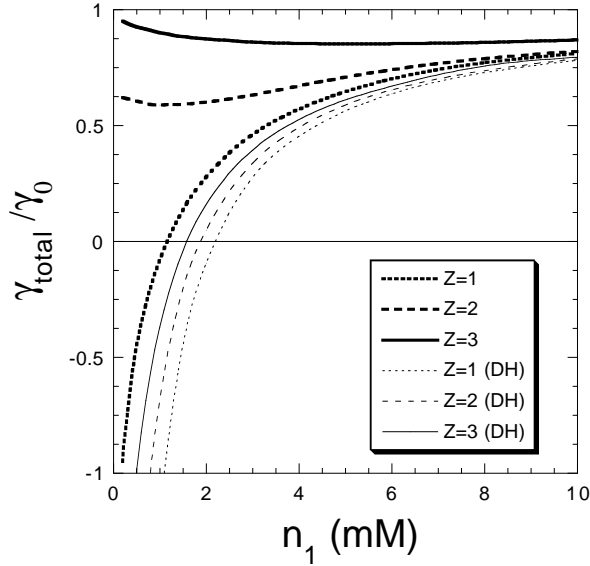}}
\caption{Total line tension, in units of the hydrophobic contribution 
$\gamma_{0}=10^{-11}$J/m, as 
a function of the monovalent salt concentration $n_{1}$.  We have chosen $T=300$ and $\sigma_{0}=0.2{\rm 
 nm}^{-2}€$.  In the absence of multivalent cations ($Z=1$), 
 there exists a finite range of the monovalent salt concentration where the membrane is unstable ($\gamma_{total}<0 $) to pore formation.  The 
 presence of as small a concentration as 0.1mM of multivalent 
 cations ($Z=2,3$) stabilizes the membrane against pore growth for 
 the whole range of  monovalent salt concentration.  The distinction 
 between the monovalent and multivalent cases is, however, minor in 
 the DH approach, and 0.1 mM of multivalent cations only 
 {\it slightly} enhances the membrane stability.     
 \label{gamma.fig}}
 \end{figure}

The charge fluctuation ($\Delta \gamma > 0$) and hydrophobic 
contributions ($\gamma_{0} >0$) favor 
closure of the pore, while the DH ($\gamma_{DH}<0$) or renormalized DH term 
($\gamma_{DH}^{*}<0$) tends to expand the  
pore.  To study the 
membrane stability, we have solved for $\sigma^{*}$ and $\gamma_{total} \equiv \gamma_{0}+ \gamma^{*} \, ({\rm 
or} \, \gamma_{DH}^{*}€)+\Delta 
\gamma$ simultaneously.  We have chosen $a_{1}=a_{2}=2 {\rm \AA}$, $T=300$, and 
$\sigma_{0}=0.2 {\rm nm}^{-2}$, which is in the physiological range.  To suppress the $R$-dependency, we have assumed that $R 
> \kappa^{-1}$.  Fig.~\ref{gamma.fig} shows 
$\gamma_{total}€$ in units of $\gamma_{0} =10^{-11}$J/m~\cite{zhelev}, as a function of 
the monovalent counterion concentration $n_{1}$.  
In the absence of multivalent counterions ($Z=1$), $\gamma_{total}€$ is
 negative when $n_{1}€$ is in the range $0 \le n_{1} \le 1{\rm mM}$.  This implies that 
the membrane is unstable to pore growth as long as $n_{1}€$ is in this range.  
The presence of 0.1mM of multivalent counterions ($Z=2$ and 3),  
however, dramatically enhances the membrane stability.  In this case, 
$\gamma_{total}€$ is positive for the whole range of $n_{1}$ and is   
in the range $0.6 \gamma_{0} \le \gamma_{total} \le 0.7 \gamma_{0}$. 
 In order to enhance the membrane stability up to this level by 
 monovalent counterions, about 5mM concentration would be needed.  This is 
approximately 50 times higher than that of the divalent counterion 
concentration.  This estimate is remarkably   
consistent with the experimental finding that ${\rm Ca}^{2+}$ is 
roughly 60 times more effective on a molar basis than ${\rm Na}^{+}$ 
in stimulating pore closure~\cite{lieber}.   

Our results are striking; the presence of multivalent counterions is more 
crucial to the membrane stability than that of the monovalent salt, 
though the ionic 
strength is mainly determined by the latter.    
This can, however, be understood in 
the context of ``counterion condensation''.  The long-ranged electrostatic 
interactions allow the membrane to adsorb
multivalent counterions {\it preferentially} even when $n_{2} \ll n_{1}$.  
These condensed multivalent counterions not only reduce the repulsion 
between backbone charges on the membrane, but they also enhance the 
strength of charge fluctuations---both of these effects are more efficient 
with multivalent cations than with monovalent ones.  
When combined, these two effects lead to significantly enhanced 
membrane stability against pore growth and vesiculation.  Also note that trivalent 
counterions are even more efficient in enhancing the membrane stability than the divalent 
counterions.  As shown in the figure, the dramatic distinction 
between the monovalent and multivalent cases is missing in the DH 
theory.   
The enhanced membrane 
stability by multivalent counterions seen in the 
experiments~\cite{steck,lieber} can be explained 
{\it only} when both the preferential adsorption of multivalent counterions and 
the effects of charge correlations are properly taken into account.

\begin{figure}
   \par\columnwidth20.5pc
   \hsize\columnwidth\global\linewidth\columnwidth
   \displaywidth\columnwidth
   \epsfxsize=3.12truein
   \centerline{\epsfbox{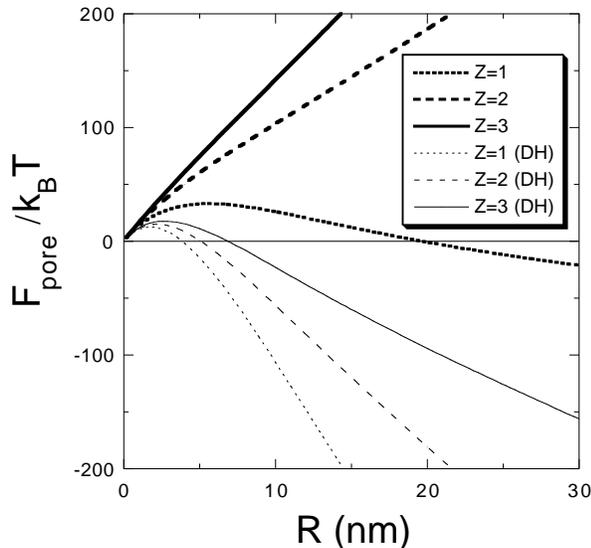}}                     
\caption{Pore free energy as a function of the pore size $R$.  We have 
chosen the same parameters as were used to generate Fig.~\ref{gamma.fig}.  The 
pore free energy is estimated in units of $k_{B}T$.  For the 
monovalent salt case ($Z=1$), the pore free energy has a finite barrier.  In 
the presence of 0.1mM of multivalent counterions ($Z=2,3$), the pore free energy 
grows {\it indefinitely} with $R$.  In the DH approach, the barrier 
height is roughly {\it in}sensitive to the counterion valency $Z$.       
\label{fpore.fig}}
\end{figure}

Whether a pore grows or closes also depends on the height of the 
barrier  as a 
function of the pore size $R$.  In Fig.~\ref{fpore.fig}, we have plotted the 
pore free energy as a function of $R$, in units of 
$k_{B}€T$.  We have chosen $T=300K$ and $\sigma_{0}=0.2{\rm 
nm}^{-2}€$.  The barrier height is finite in the presence of 
monovalent ions only ($Z=1$).  In contrast, 
the pore free energy in 
the presence of 0.1mM of multivalent counterions ($Z=2,3$) grows {\it indefinitely} with 
$R$.  This implies that formation of a large pore is energetically greatly {\it 
dis}favored, in the presence of multivalent counterions.  The 
results are indeed consistent with the experimental observation that 
the presence of 0.1mM of ${\rm MgSO}_{4}$ stabilized the red blood 
cell ghosts against vesiculation~\cite{steck}; a pore originally 
created by osmotic stress will grow into a large one or shut down, 
depending on the buffer quality and the strength of the restoring force 
provided by the spectrin network.  Note that the osmotic stress will 
eventually be removed.  When $Z=1$, the pore can grow into a 
large one, once the barrier is overcome by osmotic 
stress.  This will lead to vesiculation if the restoring 
force is outweighed by the repulsion between the charged groups on 
the red 
cell membranes.  In contrast, there is a subsequent 
barrier to pore growth in the case $Z=2,3$.  This 
prevents the ghosts from breaking into vesicles.  In contrast, the DH 
approach mistakenly implies that the barrier height is roughly 
{\it in}sensitive to the valency of counterions.  Thus our results in 
Fig.~\ref{fpore.fig} further support the importance of the 
counterion valency and charge correlations to the membrane stability, consistent 
with experiments~\cite{steck,lieber}.

The fact that multivalent cations can be preferentially adsorbed 
onto a charged surface implies that the layer of the adsorbed cations 
can be depleted by multivalent anions.  Imagine an anion of valency $Z_{-}€$ 
making a pair with a cation in the condensed layer and leaving into the 
solution, as illustrated in Fig.~\ref{model.fig}(b).  Whether this 
is feasible can be tested by calculating the change in the 
chemical potential: $\Delta \mu \sim  -Z_{+} Z_{-} \ell_{B} 
/\left(a_{2+} + a_{2-}  \right) + Z_{+} 
\ell_{B} \sigma^{*} \kappa^{-1}€+ \ln \left( n_{2-} a_{2-}^{3} /n_{2+} 
a_{2+}^{3} \right)$, where the subscript 2 refers to multivalent ions and the 
subscripts $+$ and $-$ refer to the cations and anions, respectively.  When the valency of anions is sufficiently high, this change 
can be negative.  This implies that a certain fraction of the cations in the 
layer make pairs with multivalent anions and will return to the 
solution to maintain ``chemical equilibrium''.  Note that a minimum 
concentration of multivalent anions is 
required for this process to occur.  This can be readily seen by 
taking the limit $n_{2-}\rightarrow 0$ and noting that, in this limit, the 
entropic penalty for pairing is too large.  Since only a very small  
concentration of multivalent cations is needed to enhance the membrane 
stability, the presence of an equally small content of multivalent 
anions suffices to deplete the condensed layer, effectively reducing 
the membrane stability. Cells contain  multivalent cations (e.g. 
${\rm Ca}^{2+}€$ and ${\rm Mg}^{2+}€$) as well as multivalent anions 
(e.g., ${\rm PO}^{3-}_{4}€$ and anionic proteins) and thus could 
regulate their stability in this way.  This may provide a new insight 
into the biological phenomena of ``breakdown'' and ``reassembly'' of a 
nuclear envelope during cell division.  During cell division, a nuclear envelope disintegrates into 
vesicles, which eventually reassemble into daughter 
cells~\cite{albert}.  This is suggestive of the cyclic change in the 
membrane stability: stable $\rightarrow$ unstable $\rightarrow$ 
stable.  The stabilization/destabilization of membranes by 
multivalent ions could be relevant to this flow of 
membrane stability in a dividing cell.  This is, however, complicated 
by various membrane-bounded proteins~\cite{albert} and further 
consideration is certainly warranted.  

 To summarize, we have presented a working mechanism for the 
stabilization/destabilization of membranes by multivalent ions.    
 Multivalent cations preferentially adsorbed onto a charged membrane not only 
reduce the surface charge density of the membrane but also induce a repulsive 
barrier to pore growth that favors the closure of a pore.  The main advantage of the membrane 
stabilization by multivalent cations lies in that this can be easily reversed; 
the addition of a 
small concentration of multivalent anions can reverse the membrane 
stabilization.  Our results are also consistent with the 
experimental findings that membrane adhesion by multivalent cations 
does not necessarily lead to membrane fusion and that membrane fusion 
(in the absence of fusion peptides) should be followed by lateral phase separation of the lipid molecules into 
two distinct phases~\cite{jacob}: anionic lipid-poor phases and 
anionic lipid-rich phases ``coated'' with multivalent cations.  It is the 
{\it un}coated phases that undergo a topological change and eventually 
fuse into each other, consistent with our picture of multivalent cations as 
efficient agents for stabilizing membranes against rupture.  

                            

We have benefited from useful discussions with D. Boal, M. Wortis, S. 
Davies, A. Rutenberg, M. Howard, K. Delaney, and G. Tibbits.  
  This work was 
supported by the Natural Sciences and Engineering Research Council of Canada.

 \end{multicols}

\begin{references}
\bibitem{remark} More realistically, a cell membrane is curved with 
unbalanced ionic strength between the inside and outside of the 
membrane.  We consider the stage where the osmotic stress has been 
removed via ion transport through the pore.
\end{references}
\end{document}